\documentclass{article}

\PassOptionsToPackage{numbers}{natbib}
\usepackage[preprint]{neurips_2021}

\usepackage[utf8]{inputenc} 
\usepackage[T1]{fontenc}    
\usepackage{booktabs}       
\usepackage{amsfonts}       
\usepackage{nicefrac}       
\usepackage{microtype}      
\usepackage{amsmath}
\usepackage{graphicx}
\usepackage{tipa}
\usepackage{multirow}
\usepackage{longtable}

\usepackage{subcaption}
\usepackage{color}
\usepackage[table]{xcolor}
\usepackage{threeparttable}
\usepackage{diagbox}
\usepackage{enumitem}
\usepackage{tablefootnote}
\usepackage{hyperref}
\usepackage{url}            
\usepackage{xurl}
\usepackage{breakurl}

\usepackage[edges]{forest}
\usepackage{tikz}
\usepackage{tikz-qtree}
\usetikzlibrary{fadings}
\usetikzlibrary{mindmap,trees}
\usetikzlibrary{arrows,automata,shapes,positioning,shadows,trees}
\usetikzlibrary{shapes,snakes,shadows}
\usetikzlibrary{shapes.arrows}
\usetikzlibrary{calc,shapes, positioning}
\usetikzlibrary{decorations.text}
\usetikzlibrary{matrix,chains,positioning,decorations.pathreplacing,arrows}
\usetikzlibrary{bayesnet}
\usetikzlibrary{shadows.blur}
\usetikzlibrary{shapes.geometric}

\tikzstyle{edge}=[-latex',draw=black!90,shorten <=1pt,shorten >=1pt]
\tikzstyle{redge}=[latex'-,draw=black!90,shorten <=1pt,shorten >=1pt]
\tikzstyle{dedge}=[latex'-latex',draw=black!90,shorten <=1pt,shorten >=1pt]
\tikzstyle{block}=[draw, text width=5em,align=center,shape=rectangle, rounded corners, , align=center]
\tikzstyle{nobox}=[align=center]

\definecolor{emb}{RGB}{209,228,252}
\definecolor{hidden-blue}{RGB}{194,232,247}
\definecolor{hidden-orange}{RGB}{243,202,120}
\definecolor{hidden-yellow}{RGB}{242,244,193}
\definecolor{output-purple}{RGB}{219,203,231}
\definecolor{output-green}{RGB}{204,231,207}
\definecolor{hiddendraw}{RGB}{205, 44, 36}

\tikzstyle{mybox}=[
    rectangle,
    draw=hiddendraw,
    rounded corners,
    text opacity=1,
    minimum height=1.5em,
    minimum width=5em,
    inner sep=2pt,
    align=center,
    fill opacity=.2,
    ]

\tikzstyle{emb-purple}=[
    rectangle,
    draw=output-purple!50!purple,
    fill=output-purple,
    text opacity=1,
    minimum height=1.5em,
    minimum width=1.5em,
    inner sep=0pt,
    align=center,
    fill opacity=.9,
    ]

\tikzstyle{emb-blue}=[
    rectangle,
    draw=emb!50!blue,
    fill=emb,
    text opacity=1,
    minimum height=1.5em,
    minimum width=1.5em,
    inner sep=0pt,
    align=center,
    fill opacity=.9,
]

\definecolor{colone}{RGB}{178, 34, 34}
\definecolor{coltwo}{RGB}{106, 90, 205}
\definecolor{colthree}{RGB}{255, 250, 205}
\definecolor{colfour}{RGB}{0, 139, 69}
\definecolor{colfive}{RGB}{245,238,197}
\definecolor{colsix}{RGB}{243,235,179}
\definecolor{colseven}{RGB}{241,231,163}
\title{A Survey on Image Quality Assessment}

\author{
Lanjiang Wang\\

University of Electronic Science and Technology of China \\
}

\begin{document}
\maketitle

\begin{abstract}

 Image quality assessment(IQA) is of increasing importance for image-based applications. Its purpose is to establish a model that can replace humans for accurately evaluating image quality. According to whether the reference image is complete and available, image quality evaluation can be divided into three categories: full-reference(FR), reduced-reference(RR), and non-reference(NR) image quality assessment. Due to the vigorous development of deep learning and the widespread attention of researchers, several non-reference image quality assessment methods based on deep learning have been proposed in recent years, and some have exceeded the performance of reduced -reference or even full-reference image quality assessment models. This article will review the concepts and metrics of image quality assessment and also video quality assessment, briefly introduce some methods of full-reference and semi-reference image quality assessment, and focus on the non-reference image quality assessment methods based on deep learning. Then introduce the commonly used synthetic database and real-world database. Finally, summarize and present challenges.

\end{abstract}

\section{introduction}
\label{inrtoduction}
In an era where visual intelligence products are so widespread today, images occupy a dominant position in the traffic carrier.Image based applications can be found everywhere in our daily life.Image dehazing/deraining proposed by \cite{wei2020single,li2021single,wei2021non,li2020region,wu2020subjective,li2020haze,wu2020unified,luo2020single,wu2019beyond} can be important in autopilot of smart cars, object detection methods proposed by \cite{li2018incremental,chen2020high,chen2021bal,qiu2019a2rmnet,ding2020human,li2019simultaneously,li2019headnet,qiu2020hierarchical,qiu2020offset,li2020codan,wang2020multi} can be used in monitoring of transportation hubs and image segmentation methods \cite{yang2020new,shang2019instance,guo2020deep,yang2019new,shi2020query,xu2019bounding,yang2020learning,yang2020mono,shi2018key,meng2019new,luo2018weakly} can be applied in medical imaging.However images are transmitted with varying degrees of degradation in the quality of the images eventually received by the observers - mostly humans - at the receiving end due to the hardware and software conditions of the transmission channel or the receiving device, as well as lossy compression, e.g. image JPEG compression can cause blurring and ringing effects. In this case, Image Quality Assessment (IQA) was created in order to maintain and improve the quality of the images at the receiving end. IQA is of great interest for various applications in all stages of computer image processing. For example, it can be used for image acquisition~\cite{Feng2005Psy}, image fusion~\cite{1247209}, face recognition~\cite{ji2012fusion}, and medical images~\cite{chow2016review}. IQA methods can be generally classified as subjective and objective~\cite{wang2008hvs}. Subjective methods are measured by the final recipient of all media - humans - and are the most accurate and reliable methods. However, subjective IQA requires a lot of human and material consumption, does not meet the requirements of real-time IQA systems, and appears impractical in practical applications. Therefore, the researchers propose an objective method that simulates human assessment of picture quality.

Objective image quality assessment is an important area of image processing research, which  automatically predicts image quality by means of mathematical models designed to approximate human prediction. Objective image quality assessment can be divided into three categories according to whether the reference image is complete and available: full-reference, reduced-reference and no-reference image quality assessment. Full-reference IQA methods use the entire undistorted reference image to compare with the distorted image and measure the difference between them, while reduced-reference IQA methods use part of the information in the reference image, such as the extraction of handicraft features. However, in practical applications, the reference image is difficult to obtain, making the above two methods inapplicable. Therefore, it is becoming increasingly important to develop effective NR-IQA methods. This paper will focus on the development of NR-IQA methods.

The rest of this paper is structed as follows: Section 2 provides an introduction to the concept of image and video quality assessment; Section 3 focuses on the NR-IQA method; Section 4 introduces the datasets commonly used in IQA; and finally Section 5 concludes this paper and provides an outlook.

\section{IQA}
\label{IQA}

\subsection{Definition}
\label{definition}
Factors affecting image quality come from several sources, such as brightness, contrast, composition, noise and so on. The degree of distortion caused by the above factors varies for different images, so it is difficult to determine which factor plays a major role. According to \cite{he2014objective} image quality is defined as consisting of three factors, namely fidelity, perception and aesthetics. Fidelity is the accuracy of the distorted image relative to the original image; perception is derived from the human visual system and such metrics consider visual attention \cite{borji2012state}, contrast masking \cite{klein1997seven}, etc.; and aesthetics is subjective and may include \cite{he2014objective} visual constancy, visual attention and visual fatigue.

\subsection{Classification}
\label{classification}
Image quality assessment can be divided into subjective quality evaluation and objective quality evaluation depending on the subject of the prediction. Subjective quality assessment is the most reliable method for assessing image quality, as human observers are in most cases the ultimate recipients of the image transmission system \cite{mohammadi2014subjective}. Mean opinion score (MOS) is a metric for subjective quality assessment that requires a large number of observers to evaluate. MOS is considered to be the best image quality metrics. Subjective quality assessment is a traditional method of measuring image quality and requires a group of subjects who must rate the quality of the image in a controlled testing environment. Each  individual's rating may be different. The results are finalized by processing a weighted average of each  individual's results. It can provide accurate results, but is slow in practical applications and expensive to use in practice \cite{wang2011applications}.

Objective image quality assessment can be performed automatically and accurately by means of mathematical models designed to fit the human evaluation of the input image, which can save a lot of human and material resources. Further, objective image quality assessment can be classified into three: full-reference, reduced-reference and no-reference image quality evaluation.

Full-reference image quality assessment (FR-IQA) can use the entire undistorted image and obtain an image quality score for the distorted image based on the difference between the distorted image and the original image. Figure 1 shows the flowchart for full-reference image quality assessment. Efforts to investigate full-reference image quality assessment include \cite{wang2004image,wang2003multiscale,sheikh2006image,larson2010most,zhang2011fsim}.

\begin{figure}[htbp]
    \centering
    \includegraphics{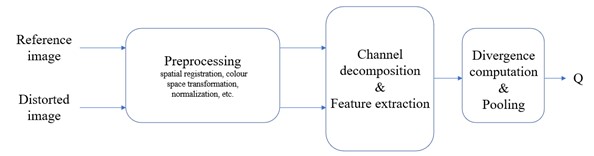}
    \caption{Framework of FR-IQA}
    \label{1}
\end{figure}

Reduced-reference image quality assessment (RR-IQA) is designed to help evaluate the quality of distorted images using only manually extracted features of the reference image when the reference image is not completely available. The idea of reduced-reference image quality assessment was originally proposed in the context of tracking the degree of visual quality degradation of video transmitted over communication networks \cite{he2014objective}. The framework of reduced-referential image quality evaluation is shown in Figure 2. Efforts to study semi-referential image quality evaluation include \cite{rehman2012reduced,wang2005reduced,li2009reduced,gu2013new,wu2016orientation}.

\begin{figure}[htbp]
    \centering
    \includegraphics{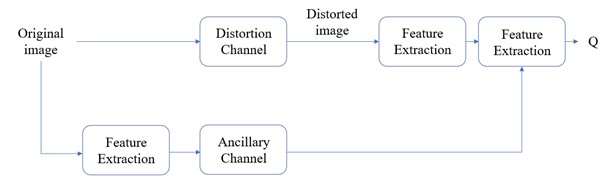}
    \caption{Framework of RR-IQA}
    \label{2}
\end{figure} 

In most practical applications it is difficult to obtain information about the reference image and often the receiver only receives a distorted image. For example, when we take a picture with a mobile phone, we can only obtain a picture with various possible truths, but not a reference image. Therefore reference-free image quality assessment (NR-IQA), also known as blind image quality assessment (BIQA), seems so important. In fact, this is a very difficult task, about which we will focus in the next section. Figure 3 shows the basic framework for most of the non-referenced image quality evaluations.

\begin{figure}[htbp]
    \centering
    \includegraphics{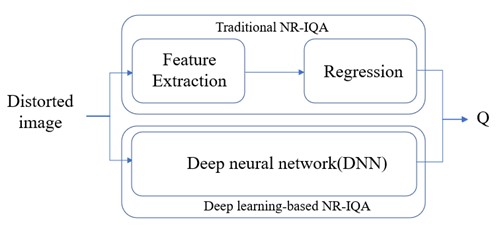}
    \caption{Framework of NR-IQA}
    \label{3}
\end{figure} 

\subsection{VQA}
\label{VQA}
Video quality assessment (VQA) aims to build models for evaluating video quality in the streaming and social media industries. Same as IQA, VQA can be divided into full-reference (FR),reduced-reference (RR) and no-reference(NR) VQA. While (FR) VQA research is maturing and several algorithms\cite{wang2004image,li2016toward} are widely used, more recent attention has turned to creating better no-reference (NR) VQA models that can be used to predict and monitor the quality of synthetically and authentically distorted  videos. Many researchers have investigated and proposed possible solutions to the NR VQA problem \cite{mittal2012no,saad2014blind,xu2014no,mittal2015completely,ghadiyaram2017perceptual,varga2019no,li2019quality}, and one simple but reasonably effective strategy is to compute quality scores for each frame and then represent the evolution or relative importance over time by applying a time pool to the quality scores. Simple time-averaged pooling is widely used for enhancing FR \cite{seshadrinathan2009motion,bampis2017speed,vu2014vis3} and NR VQA models\cite{saad2014blind,mittal2015completely,varga2019no}. Other types of pooling used include harmonic averaging \cite{li2018vmaf}, Minkowski averaging \cite{rimac2009influence,seufert2013pool}, percentile pooling \cite{moorthy2009visual,chen2016perceptual} and adaptive weighted sums \cite{park2012video}. More sophisticated pooling strategies consider memory effects such as prioritization, recency \cite{rimac2009influence,seufert2013pool,bampis2017study} and hysteresis\cite{xu2014no,li2019quality,seshadrinathan2011temporal,choi2018video}. However, to date, the general applicability of these pooling models has not been thoroughly validated in the general context of real-world UGC video NR VQA models, although some more focused research has been conducted \cite{rimac2009influence,seufert2013pool}.

\section{No-reference IQA}
\label{NR-IQA}

No-reference image quality assessment (NR-IQA), also known as blind image quality assessment (BIQA), aims to construct a computational model for accurately and automatically predicting the quality of images as perceived by humans without the need for additional information. As can be seen in Figure 3, the first type of NR-IQA methods are based on traditional regression methods, where manually designed features are first extracted from the image and then a trained regression network is used to predict the image quality to obtain the final quality score. Earlier studies used a priori knowledge for predicting specific distortion types \cite{li2015no,liu2009no} by extracting and exploiting specific distortion features. Li et al \cite{liu2009no} proposed an NR-IQA method based on fuzzy distortion. They used a gradient image to describe the ambiguity and segmented the gradient image into blocks, extracting the energy features of each block associated with the ambiguity distortion. Finally, the image quality is obtained by normalization. The disadvantage of this type of method is that it is difficult to find efficient features for quality assessment if the image shows distortions that have not been manually designed.

To address the problem of assessing image quality without a distortion prior, researchers have used natural statistics (NSS)-based methods to extract reliable features which assume that natural images share certain statistical information and that distortion may alter these statistics \cite{wu2015highly,zhang2015feature,wu2015blind,yang2018blind,ji2019blind,wu2017blind,zhou2019no}.Wu et al. select statistical features extracted from binary patterns of local image structures. Zhang et al. used multivariate Gaussian (MVG) models, Yang et al. used generalized Gaussian distribution ( GGD) model to extract image features, and Wu et al. used multi-channel fused image features to simulate the hierarchical and trichromatic properties of human vision. Then, k-nearest neighbor (KNN) based models were used to evaluate image quality. The following table shows the various representative evaluation models. The feature extraction methods, regression methods and datasets used by the three representative IQA models are listed in the table.

\begin{table}[h!]
  \begin{center}
    \caption{Comparison of representative models.}
    \begin{tabular}{l|c|c|r}
      \textbf{Method} & \textbf{Feature Extraction} & \textbf{Regression} & \textbf{Databases}\\
      \hline
      IL-NIQE\cite{zhang2015feature} & MVG & Pooling & TID2013,CSIQ, LIVE, MD1,MD2\\
      BWS\cite{yang2018blind} & GGD & SVR & LIVE, TID2008, CSIQ\\
      TCLT\cite{wu2015blind} & Multichannel Feature Fusion & KNN &LIVE II,
CSIQ,TID2008
\\
    \end{tabular}
  \end{center}
\end{table}

However, these NR-IQA methods are limited to hand-crafted features that may not adequately represent complex image structures and distortions. The rise of deep learning in recent years has piqued the interest of researchers. Deep neural networks (DNNs) can automatically extract deep features relevant to quality assessment and optimize these features by using back-propagation methods to improve predictive performance \cite{yang2019survey}. As a result, DNNs can be applied to various image quality assessment (IQA) methods such as \cite{wu2019blind,meng2019new} and becomes a promising option for solving NR-IQA tasks. It is well known that deep learning techniques have been highly successful in solving various image recognition and target detection tasks \cite{krizhevsky2012imagenet,szegedy2015going,he2016deep,simonyan2014very}. The main reason is that it relies heavily on large-scale annotated data, such as the ImageNet dataset, which is oriented towards image recognition. However, for NR-IQA tasks, due to the relatively small size of the dataset, direct training with DNNs leads to overfitting, i.e., the trained model has perfect performance for the training set, but unreliable performance for the tested data. Researchers have therefore focused on exploring effective DNN-based NR-IQA methods to address this problem. Some of the prominent work is presented below.

\subsection{DeepRN}
\label{DeepRN}
Varga et al. \cite{varga2018deeprn}propose DeepRN, a blind image quality assessment (BIQA) method based on deep learning of convolutional neural networks (CNN). DeepRN uses a residual deep learning network (ResNet-101) as the backbone network to extract features and adds an adaptive spatial pyramidal pooling layer at the end of the network so that the backbone network can output fixed size features, so that images of arbitrary size can be processed. Training was performed on a new, large, labelled dataset of 10073 images (KonIQ-10k) that contained histograms of quality ratings in addition to mean opinion scores (MOS).DeepRN obtained the leading performance at the time.


\subsection{TS(Two-Stream) model}
\label{TS(Two-Stream) model}

Yan et al. \cite{yan2018two}propose a two-stream convolutional network that takes an image and a gradient map as input respectively thereby acquiring different levels of information from the input and alleviating the difficulty of extracting features from a single input. The distorted local non-uniform distribution in the image is also considered by adding a region-based full convolution layer for using the information around the center of the input image block. The final score for the overall image is calculated by averaging the block scores. Experimental results on a range of benchmark datasets, such as LIVE, CISQ, IVC, TID2013, and Waterloo Exploration datasets, show that the dual-stream network performs reliably. 


\subsection{RankIQA}
\label{RankIQA}

Liu et al. \cite{liu2017rankiqa}proposed RankIQA , an innovative application of ranked training to learn image quality assessment criteria. To address the problem of limited dataset size, RankIQA uses a twin network (Siamese Network) trained on a large synthetic dataset (using synthetic distortions of known relative image quality, which can be generated automatically without manual annotation) to rank distortion levels. Once trained, a branch of the network is used to fine-tune it on the IQA dataset, mapping the features to the annotated values. Experiments on the TID2013 benchmark showed that RankIQA improved performance by over 5\%. There is also a lot of work on NR-IQA that borrows the idea of Learning-to-Rank, such as DipIQ \cite{ma2017dipiq} proposed by Ma et al. and PieAPP \cite{prashnani2018pieapp} proposed by Prashnani et al. Both have achieved good results by learning to rank image pairs to obtain an image quality-aware model.


\subsection{Hallucinated-IQA}
\label{Hallucinated-IQA}

Lin et al \cite{lin2018hallucinated} applied Generative Adversarial Networks (GAN) to NR-IQA to solve the NR-IQA problem from a novel perspective. They first generated an illusionary reference image from the distorted image to compensate for the missing real reference image, and then fed the difference between the distorted image and the illusionary reference to a regression network to generate quality predictions. The network consists of three components, including a quality-aware generation network G, a specially designed judgement network D and a quality regression network R. The training strategy is to first train the GAN network to generate a large number of illusionary images, which are similar to the reference images in the IQA database. Then, the R network is trained to predict the image quality scores. In the GAN network, the D network is first trained to distinguish between the pseudo-reference images in the IQA database and the reference images. Then, the G network is trained to generate images that are similar to the real reference images in the IQA database. Finally, the image quality score can be predicted by optimizing the loss of the R network. The model gives good results on several artificial, real distortion datasets.


\subsection{TRIQ}
\label{TRIQ}

First proposed in 2017 and replacing RNNs in the field of natural language processing (NLP), Transformer has also attracted research interest in the field of computer vision \cite{carion2020end,DBLP:journals/corr/abs-2010-11929}.You et al \cite{DBLP:journals/corr/abs-2101-01097} investigated the application of Transformer in image quality (TRIQ) evaluation in the visual Transformer (ViT) the Transformer encoder uses adaptive positional embedding techniques to process images of arbitrary resolution. The model was trained using the large datasets LIVE-wild \cite{ghadiyaram2015massive} and KonIQ-10k \cite{lin2018koniq} Different settings of the Transformer architecture were tried on publicly available image quality databases and the results showed that the TRIQ architecture achieved excellent performance. 


\section{Datasets and Performance Metrics}
\label{Datasets and Performance Metrics}

Datasets and performance metrics are used to evaluate the performance of various algorithms. The datasets and performance metrics commonly used in IQA are described as follows.

\subsection{Datasets}
\label{Datasets}

he most used datasets in the IQA are LIVE \cite{sheikh2006statistical}, TID2008 \cite{ponomarenko2009tid2008}, TID2013 \cite{ponomarenko2013color}, CSIQ \cite{larson2010most}, LIVE MD \cite{jayaraman2012objective}, LIVE In the Wild Image Quality Challenge Database \cite{ghadiyaram2015massive}, KonIQ-10k \cite{lin2018koniq}, etc. Among them, LIVE In the Wild Image Quality Challenge Database and KonIQ-10k are natural distortion datasets, and the rest five are artificially simulated distortion datasets.

\subsubsection{LIVE}
\label{LIVE}

The LIVE (Laboratory for Image \& Video Engineering) dataset  \cite{sheikh2006statistical} was created by the LIVE Laboratory at the University of Texas at Austin and is one of the most used datasets for IQA. The dataset contains 29 reference images with resolutions ranging from 438 × 634 pixels to 512 × 768 pixels. These images were manually simulated to produce 5 different types of distorted images, 779 in total. Distortion types include JPEG2000 distortion, JPEG distortion, white noise distortion, Gaussian blur distortion and fast Rayleigh decay distortion. The dataset provides the Differential Mean Opinion Score (DMOS) for all distorted images in the range [0, 100], where 0 means no distortion.

\subsubsection{TID2008}
\label{TID2008}

TID ( Tampere Image Database) 2008 \cite{ponomarenko2009tid2008} was created by Tampere University of Technology, Finland in 2008 and includes 25 color reference images with a resolution of 384 × 512 pixels, with 17 distortion types, each containing four different levels of distortion, for a total of 1700 images. Artificial distortions include additive Gaussian noise, spatial correlation noise, masking noise, high frequency noise, impulse noise, quantization noise, Gaussian blur, image denoising, JPEG compression, JPEG2000 compression, JPEG transmission error, JPEG2000 transmission error, non-offset pattern noise, varying intensity of local block-by-block distortion, average offset (intensity shift), and contrast variation. The data set provides MOS values and their standard deviations for all tested images, with MOS values in the range [0, 9], where 9 means that the image is distortion-free.

\subsubsection{TID20123}
\label{TID2013}

TID2013 \cite{ponomarenko2013color} was created by Tampere University of Technology, Finland in 2013 and includes 25 reference images, 3000 distorted images (including 25 reference images with 24 distortions simulated manually for each image, each with 5 levels of distortion). There are 24 distortion types, an increase of 8 distortion types compared to the TID2008 image dataset, namely altered color saturation index, lossy compression, multiple Gaussian noise, color image quantization, sparse sampling, chromatic aberration and comfort noise. The annotation type for this dataset is DMOS, which is obtained statistically from 524,340 samples observed by 971 observers, with DMOS values ranging from [0, 9], with larger values indicating poorer image quality.

\subsubsection{CSIQ}
\label{CSIQ}

CSIQ (Categorical Subjective Image Quality) \cite{larson2010most} was established by Oklahoma State University, USA, which contains 30 reference images with a resolution of 512 pixels × 512 pixels and six types of distortion, including overall contrast reduction, JPEG compression, JPEG2000 compression, additive Gaussian pink noise, additive Gaussian white noise and Gaussian blur. Additive Gaussian white noise and Gaussian blur, each containing 4 to 5 distortion levels, for a total of 866 distorted images. The data set also provides the DMOS values for all the images tested, obtained from several subjective ratings by 25 testers, with the DMOS values ranging from [0, 1], with higher values indicating poorer image quality.

\subsubsection{LIVE MD}
\label{LIVE MD}

The LIVE MD database \cite{jayaraman2012objective} is the first database containing multiple distorted images created by the LIVE Lab at the University of Texas at Austin, USA. The images are combined into two types of distortion: JPEG compression and Gaussian blur, then white noise distortion is added. It contains 15 reference and 450 distorted images and provides a DMO for each distorted image, taking values in the range [0, 100].

\subsubsection{LIVE In the Wild Image Quality Challenge Database}
\label{LIVE WILD}

This is the LIVE dataset \cite{ghadiyaram2015massive} of 1162 images from the Field Image Quality Challenge, created by the LIVE Lab at the University of Texas at Austin, USA. The types of distortion in these images are not computer simulations, but real images captured using various mobile cameras such as smartphones and tablets. During the imaging process, these images are subject to a variety of randomly occurring distortions and real photography artefacts. In order to ensure that the subjective quality scores of these images are objective, the researchers designed and implemented a new online crowdsourcing system in which 8100 observers conducted IQA and obtained 350,000 opinion scores, which were combined to produce an evaluation result.

\subsubsection{KonIQ-10k}
\label{K-10k}
KonIQ-10k \cite{lin2018koniq} was built at the University of Konstanz, Germany, to address the problem of too small the true distortion datasets. KonIQ-10k randomly selected approximately 10 million images from the large public multimedia dataset YFCC 100M \cite{thomee2016yfcc100m} and then filtered 10,073 images in stages for use in building the dataset. The types of distortion present in these images included noise, JPEG artefacts, blending, lens motion blur, over-sharpening, etc. Based on the collected dataset, the researchers conducted a large-scale crowdsourcing experiment to obtain 1.2 million evaluations from 1467 workers, using statistical methods such as taking the mean value and removing extreme scores to determine the final MOS value. The image size was 1024 pixels by 768 pixels, and the MOS values ranged from [0, 5], with higher values indicating less distortion.

\subsection{Performance Metrics}
\label{PM}
IQA has a number of performance metrics. PLCC, SROCC, KROCC and RMSE are by far the most used, as defined below.
\subsubsection{PLCC(Pearson Linear Correlation Coefficient)}
\label{PLCC}
$$PLCC(Q_{est},Q_{sub})=\frac{cov(Q_{sub},Q_{est}}{\sigma(Q_{sub})\sigma(Q_{est})}$$
where $Q_{est}$ and $Q_{sub}$ are the sets of predicted and actual subjective scores, respectively, $cov(.)$ denotes the covariance between $Q_{est}$ and $Q_{sub}$, and $\sigma(.)$means standard deviation. PLCC describes the correlation between the result of the algorithm and the subjective scoring by the human eye, and can measure the accuracy of the algorithm.

In this review, we paid attention to the work on audio synthesis and audio-visual multimodal processing. The text-to-speech(TTS) and music generation tasks, which plays a crucial role in the maintenance of the audio synthesis field, were comprehensively summarized respectively. In the TTS task, two-stage and end-to-end methods were distinguished and introduced. As for the music generation task, symbolic domain and raw audio domain generative models were presented respectively. In the field of audio-visual multimodal processing, we focused on four typical tasks: lipreading, audio-visual speech separation, talking face generation and sound generation from video. The frameworks related to these tasks were introduced. Finally, several widely adopted datasets were also presented. Overall, this review provides considerable guidance to relevant researchers.
\subsubsection{SROCC(Spearman Rank-Ordered Correlation Coeffcient)}
\label{SROCC}
$$SROCC(Q_{est},Q_{sub})=1-\frac{6\sum(d_i)}{m(m^2-1)}$$
where $d_i$ is the grade difference between the $i_{th}$ sample of $Q_{est}$ and $Q_{sub}$, and $m$ is the number of images of the surrogate evaluation database species. The SROCC is primarily used to measure the monotonicity of the algorithm's predictions.

\subsubsection{KROCC(Kendall Rank Order Correlation Coeffcient)}
\label{KROCC}
$$KROCC=\frac{2n_c-n_d}{n(n-1)}$$
Where $n_c$ the number of consistent element pairs in the dataset; $n_d$ is the number of inconsistent element pairs in the dataset. KROCC is also a good measure of the monotonicity of the algorithm.
\subsubsection{RMSE(Root Mean Squared Error)}
\label{RMSE}
$$RMSE=[{\frac{1}{n}\sum_{i=1}^n(x_i-y_i)^2}]^\frac{1}{2}$$
where $x_i$ is the subjective MOS value and $y_i$ is the quality prediction score. RMSE is a direct measure of the absolute deviation between a person's subjective score and the algorithm's predicted score.

\section{Conclusion and Overlook}
\label{Conclusion and Overlook}
This paper summarizes and reviews the basic concepts and classifications in the field of IQA, lists representative methods for each type of application, and focuses on the NR-IQA approach. NR-IQA has now become the focus of research. NR-IQA has a wide range of applications and is of significant research value, but it is also more difficult to study and requires more problems to be solved. The existing methods perform well on synthetic distortion datasets, but do not achieve good results on more challenging datasets such as LIVE Wild, KonIQ-10k and other real distortion datasets. As the distortion types are complex in most practical applications, it is difficult to generate corresponding distortion images using computer simulations, so there is more scope for realistic distortion-oriented IQA applications. There is still much scope for research into this type of evaluation task. In future research work, it is possible to expand the data set and design IQA models that are more robust to realistic distortion; to create performance metrics that are more consistent with the visual properties of the human eye, to make the evaluation results consistent with human subjectivity, etc.

\bibliography{main}
\bibliographystyle{plainnat}

\end{document}